 \documentstyle[preprint,tighten,prc,aps]{revtex}
\begin{document}
\draft
\preprint{\vbox{. \hfill ADP-99-4-T347}}

\title{Strange hadron matter and SU(3) symmetry}
\author{V.G.J. Stoks$^{1,2}$ and T.-S. H. Lee$^{1}$}
\address{$^1$ Physics Division, Argonne National Laboratory,
         Argonne, Illinois 60439    \protect\\
         $^2$ Centre for the Subatomic Structure of Matter,
         University of Adelaide, SA 5005, Australia}
\date{}
\maketitle

\begin{abstract}
We calculate saturation curves for strange hadron matter using
recently constructed baryon-baryon potentials which are constrained
by SU(3) symmetry. All possible interaction channels within the
baryon octet (consisting of $N$, $\Lambda$, $\Sigma$, and $\Xi$)
are considered. It is found that a small $\Lambda$ fraction in
nuclear matter slightly increases binding, but that larger
fractions ($>10\%$) rapidly cause a decrease. Charge-neutral
$\{N,\Lambda,\Xi\}$ systems, with equal densities for nucleons and
cascades, are only very weakly bound.
The dependence of the binding energies on the strangeness per baryon,
$f_s$, is predicted for various $\{N,\Lambda,\Xi\}$ and
$\{N,\Lambda,\Sigma,\Xi\}$ systems.
The implications of our results in relativistic heavy-ion collisions
and the core of a dense star are discussed. We also discuss the
differences between our results and previous hadron matter calculations.
\end{abstract}

\narrowtext

\section{INTRODUCTION}
\label{sec:intro}
The study of the properties of strangeness-rich systems is of
fundamental importance in understanding relativistic heavy-ion
collisions~\cite{Kum95} and some astrophysical problems~\cite{Bay95}.
The qualitative features of such systems and their possible detection
in the universe and in relativistic heavy-ion collisions
were first discussed by Bodmer~\cite{Bod71} in 1971.
Within Quantum Chromodynamics (QCD), it was suggested that the
strangeness-rich systems could be strange quark systems consisting
of up ($u$), down ($d$), and strange ($s$) quarks. These exotic systems
could be either metastable states~\cite{Chi79} against the decays
into hadrons, or absolute bound states with energies much lower
than normal nuclear matter~\cite{Wit84,Ber87}.
However, the theoretical calculations for strange quark systems are
still in the developing stage. For example, within the MIT bag model
it is found~\cite{Sch97} that the stability of the bag strongly depends
on the rather uncertain bag constant $B_{\rm bag}$. The strange quark
matter is absolutely stable for $B^{1/4}_{\rm bag}\sim140$ MeV,
metastable for $B^{1/4}_{\rm bag}=150$--$200$ MeV, and unstable for
$B^{1/4}_{\rm bag}>200$ MeV.
Therefore, one cannot rule out the possibility that the strangeness-rich
systems could be strange hadron systems made of nucleons and hyperons;
as studied, for example, in Refs.~\cite{Pan71,Sch93,Sch94,Gal95,Sch98}.

The production and detection of strangeness-rich matter from
relativistic heavy-ion collisions has been studied in recent
years~\cite{Gre87,Gre88,Gre91,Cra92,Bal94,Sch97}.
Two scenarios have been discussed. The first one is the coalescent
mechanism~\cite{Bal94}, which assumes that the produced hyperons are
captured by the nearby nuclear fragments in the freeze-out region to
form multi-hyperon clusters. The second mechanism is the distillation
process~\cite{Gre87,Gre88,Gre91,Sch97} associated with the production
of a quark-gluon plasma (QGP) in the baryon-rich region. The essential
idea is that the $\bar{s}$ quarks in a QGP, in which $s\bar{s}$ pairs
are abundant, are captured by the surrounding $u$ and $d$ quarks,
liberated from the initial heavy ions, to form $K^+$ and $K^0$.
The emission of these kaons and other mesons causes cooling of the
QGP into a strange quark system.
The interesting question to ask is whether this system will be absolutely
stable or metastable against the collapse into a hadron system.
This question can be answered by comparing the energy of such a system
with the energy of a strange hadron system with the same strangeness
quantum number. It is therefore important to develop theoretical
approaches to calculate the energies of strange hadron systems.
In this paper, we report on the first results of our efforts in this
direction. The calculation for the strange quark systems with similar
sophistication is beyond the scope of this paper.

Most of the recent investigations of strange hadron systems
have been done by using the relativistic mean-field
model~\cite{Sch93,Sch94,Gal95,Sch97}. In addition to the usual
$\sigma$ and $\omega$ mesons, these models also contain $\sigma^*$
and $\phi$ mesons, introduced in order to have strong attractive
hyperon-hyperon interactions. The vector coupling constants are chosen
according to SU(6) symmetry, while the scalar coupling constants are
fixed to hypernuclear data. Extensive calculations for the systems
consisting of $\{p,n,\Lambda,\Xi^0,\Xi^-\}$ mixtures have been performed.
Clearly, these calculations are not completely consistent with the
SU(3) symmetry, since $\Sigma$'s are not included (for reasons which
will be discussed below).
Furthermore, it is not clear that the values for the coupling constants
employed in these models are consistent with the very extensive data
on nucleon-nucleon ($N\!N$) and hyperon-nucleon ($Y\!N$) reactions.
A rigorous prediction should be consistent with both the two-body data
and the data of hypernuclei.

An alternative approach is based on the many-body theory with
baryon-baryon potential models. This was first pursued by
Pandharipande~\cite{Pan71} using a variational method and rather
crude baryon-baryon potentials.
This approach has recently be revived in Ref.~\cite{Sch98} by using
the Brueckner-Hartree-Fock approximation and the Nijmegen soft-core
$Y\!N$ potential~\cite{Mae89} of 1989. There, the authors only consider
an infinite system consisting of $\Lambda$'s and nucleons.
It was found that for a given total baryon density $\rho_B$ the binding
energy per baryon, $E_B=-E/A_B$, decreases as the fraction of
strangeness, $f_s=|S/A_B|$, increases.
The most stable system in their approach has $E_B=-E/A_B\sim16.0$ MeV,
at $f_s\sim0.06$ and $\rho\sim0.23$ fm$^{-3}$.
Comparing with the calculations using relativistic mean-field models
as described above, this investigation is rather incomplete since the
role of $\Xi$ is not explored, owing to the restriction of the employed
Nijmegen potential. The importance of $\Xi$ was pointed out in
Ref.~\cite{Sch93}. It is needed to stabilize the system against the
strong $\Lambda+\Lambda\rightarrow \Xi+N$ process, which can occur
at relatively low density of $\Lambda$ since the threshold energy for
this reaction to occur is only about 28 MeV. The presence of $\Xi$
will Pauli block this reaction.
Furthermore, their calculations do not include hyperon-hyperon
interactions and hence the effects of additional hyperons on the
hyperon mean field are neglected in solving the self-consistent
$G$-matrix equation.
In this paper we try to be as complete as possible in that we
include all possible interaction channels that are allowed for
the baryon octet. We use the recently constructed NSC97 baryon-baryon
potential models~\cite{Rij99,Sto99} to describe all these channels.

The content of the paper is as follows. In Sec.~\ref{sec:model} we
briefly highlight some of the features of the employed baryon-baryon
potential models. In Sec.~\ref{sec:theory} we review the definition
of the $G$ matrix and related quantities, with some emphasis on the
treatment of the coupled channels. In Sec.~\ref{sec:results} we present
and discuss the results of our calculations. We conclude with a
brief summary of our findings in Sec.~\ref{sec:conc}.

\section{BARYON-BARYON POTENTIAL}
\label{sec:model}
In this paper, we report on the first results from an investigation
of strange hadron matter using the most recently developed
baryon-baryon potentials~\cite{Rij99,Sto99}. These potentials are
constructed within the dynamics defined by the SU(3) symmetry, and the
data of nucleon-nucleon ($N\!N$) and hyperon-nucleon ($Y\!N$) reactions.
We follow a similar Brueckner-Hartree-Fock formulation as recently
employed by Schulze {\it et al.}~\cite{Sch98}, but with an important
improvement: the starting coupled-channel $N\!N\oplus Y\!N \oplus YY$
potentials include the $\Lambda\Lambda$, $\Lambda\Sigma$, $\Sigma\Sigma$,
and $\Xi N$ channels of strangeness $S=-2$, as required by SU(3) symmetry.
The presence of $\Xi$ then also enforces us to consider the $\Xi\Lambda$
and $\Xi\Sigma$ channels of strangeness $S=-3$, and the $\Xi\Xi$
channel of strangeness $S=-4$. The assumption of SU(3) symmetry
allows us to unambiguously define the baryon-baryon interactions for
the $S=-2,-3,-4$ systems from the previously constructed $N\!N$ and
$Y\!N$ interactions.

However, because of the lack of sufficiently accurate $Y\!N$ data and
some uncertainties in SU(3) coupling constants, the constructed
baryon-baryon potentials have some model dependence. In Ref.~\cite{Rij99},
six $Y\!N$ models have been constructed, based on different choices for
the vector-magnetic $F/(F+D)$ ratio, $\alpha_V^m$. Values range from
$\alpha_V^m=0.4447$ for model NSC97a to $\alpha_V^m=0.3647$ for model
NSC97f. The different choices for $\alpha_V^m$ (consistent with
static or relativistic SU(6) predictions) are chosen such that the
models encompass a range of scattering lengths in the $\Sigma N$ and
$\Lambda N$ channels, but all models describe the $Y\!N$ scattering
data equally well. Differences show up in more elaborate applications
such as hypernuclear calculations; see Ref.~\cite{Rij99} for a more
detailed discussion. In this paper we will only consider models NSC97e
and NSC97f, which seem to be the most consistent with the existing
hypernuclear data~\cite{Rij99}.

Another important assumption in the construction of the baryon-baryon
potentials is that the SU(3) symmetry is applied to the full range of
the interaction; i.e., to the long-range as well as to the short-range
part. Although there is no empirical evidence that the short-range
part indeed satisfies the SU(3) symmetry (there are no $YY$ scattering
data to test this assumption, for example, with these potential models),
we have chosen for this approach since it allows us to extend the
$N\!N$ and $Y\!N$ interactions to all $YY$ interactions describing the
$S=-2,-3,-4$ systems, without having to introduce any new parameters.
To illustrate the differences between the various interactions, we show
in Fig.~\ref{strange:phs1s0} the $^1S_0$ elastic phase shifts for
the two models in the $N\!N$, $\Lambda\Lambda$, $\Sigma\Sigma(T=2)$,
and $\Xi\Xi$ channels.
The differences between NSC97e and NSC97f are fairly small and, at
this scale, will only show up in the $\Lambda\Lambda$ channel.
A more detailed description of the NSC97 potential models in the $YY$
channels will be presented elsewhere~\cite{Sto99}.

\section{$\protect\bbox{G}$-MATRIX FOR COUPLED CHANNELS}
\label{sec:theory}
In the next section we will present the results of several
Brueckner-Hartree-Fock calculations using the baryon-baryon potentials
for the $S=0,\ldots,-4$ systems, and so here we review some of the
aspects of the $G$ matrix and define the relevant quantities.
In each case, the (strange) nuclear matter is characterized by a
total density $\rho$, which is broken up into the contributions from
the four baryon species according to
\begin{eqnarray}
   \rho &=& \rho_N + \rho_\Lambda + \rho_\Sigma + \rho_\Xi       \nonumber\\
        &=& \rho(\chi_N + \chi_\Lambda + \chi_\Sigma + \chi_\Xi) \nonumber\\
        &=& \frac{1}{3\pi^2}\left(2k_F^{(N)3}+k_F^{(\Lambda)3}
                     +3k_F^{(\Sigma)3}+2k_F^{(\Xi)3}\right).
                                   \label{rhotot}
\end{eqnarray}
This also defines the Fermi momentum $k_F^{(B)}$ for a baryon $B$ with
density fraction $\chi_B$. As standard, we define the $G$ matrix
$G({\bf p}'_1,{\bf p}'_2;{\bf p}_1,{\bf p}_2)$ for incoming (unprimed)
and outgoing (primed) momenta.
Defining mass fractions $\mu_i=M_i/(M_1+M_2)$, the total momentum
is ${\bf P}={\bf p}_1+{\bf p}_2={\bf p}'_1+{\bf p}'_2$, and the
relative momenta are given by ${\bf k}=\mu_2{\bf p}_1-\mu_1{\bf p}_2$
and ${\bf k}'=\mu_2{\bf p}'_1-\mu_1{\bf p}'_2$.
The $G$ matrix satisfies the Bethe-Goldstone equation~\cite{Bru58}
\begin{equation}
   G({\bf k}',{\bf k};{\bf P},\omega) = V({\bf k}',{\bf k})
   +\int\!\!\frac{d^3q}{(2\pi)^3} V({\bf k}',{\bf q})
    \frac{Q({\bf q},{\bf P})}
   {\omega-E_1(\mu_1{\bf P}\!+\!{\bf q})-E_2(\mu_2{\bf P}\!-\!{\bf q})}
    G({\bf q},{\bf k};{\bf P},\omega),      \label{BGequ}
\end{equation}
where $Q({\bf q},{\bf P})$ is the Pauli operator which ensures that
the intermediate-state momenta are above the Fermi sea (see below).
The $\omega$ denotes the starting energy (including rest masses),
while the intermediate-state energies are given by
\begin{eqnarray}
   E_1({\bf p}''_1)+E_2({\bf p}''_2) &=& M_1\!+\!M_2
      +\frac{p_1^{\prime\prime2}}{2M_1}+\frac{p_2^{\prime\prime2}}{2M_2}
      +{\cal R}e\,U({\bf p}''_1)+{\cal R}e\,U({\bf p}''_2)    \nonumber\\
   &=& M+\frac{P^2}{2M}+\frac{q^2}{2\mu}
      +{\cal R}e\,U(\mu_1{\bf P}\!+\!{\bf q})
      +{\cal R}e\,U(\mu_2{\bf P}\!-\!{\bf q}),    \label{spener}
\end{eqnarray}
with $M$ and $\mu$ the total and reduced mass, respectively.
The single-particle potentials $U$ are defined by
\begin{equation}
   U({\bf p}_1)=\int^{(k_F^{(2)})}\!\!\frac{d^3p_2}{(2\pi)^3}
                G\left[{\bf p}_1,{\bf p}_2;{\bf p}_1,{\bf p}_2;
                \omega=E_1({\bf p}_1)+E_2({\bf p}_2)\right]. \label{sppot}
\end{equation}
Hence, we have two equations, Eqs.~(\ref{BGequ}) and (\ref{sppot}),
which have to be solved self-consistently.

In our calculations we have made several approximations. First of all,
the energies are treated in the nonrelativistic expansion, as is obvious
from the $1/M^2$ expansion in Eq.~(\ref{spener}). The same $1/M^2$
expansion was used in the derivation of the NSC97 potentials~\cite{Rij99}.
Brockmann and Machleidt~\cite{Bro90} have demonstrated the effect
on nuclear-matter results when one uses instead the relativistic
energies and the Dirac equation for the single-particle motion; these
type of calculations have become known as Dirac-Brueckner calculations.
For purely nuclear matter they find that the saturation point shifts
to lower density and has a smaller binding energy per nucleon.
In analogy with their result, we expect that a proper Dirac-Brueckner
calculation for strange nuclear matter will also show a shift of
saturation points as compared to what we find in our present
Brueckner-Hartree-Fock calculations. However, we believe that the
calculations presented here suffice for our purpose, which is to
study the general features of strange nuclear matter. Any shift
of saturation points is only expected to be relevant in those cases
where the matter under consideration is on the boundary of being
bound or unbound. In those cases, the Dirac-Brueckner result might
show that matter which we find to be just bound is actually unbound,
or vice versa. 

A second approximation is that the single-particle potential is
radically put to zero for momenta ${\bf p}_i$ above the Fermi sea:
the so-called ``standard'' choice. This causes a discontinuous jump
in $E_i(p_i)$ at $p_i=p_F$, and so it is also known as the ``gap''
choice. An alternative choice is to retain a nonzero value for momenta
above the Fermi sea: the so-called ``continuous'' choice~\cite{Jeu76}.
There are various physical arguments which favor this latter
choice~\cite{Jeu76}, but its main effect is to merely shift the
saturation curve to give more binding, without changing the overall
density dependence very much; see, e.g., Refs.~\cite{Bal91,Sch95} for
the effect in ordinary nuclear matter. However, these differences are
only of relevance on a quantitative level (e.g., when a comparison
is made with the experimental saturation point), and so we argue
that for this first study of the general features of strange nuclear
matter it suffices to use the gap choice.
Another motivation is that the continuous choice considerably
complicates the propagator in the Bethe-Goldstone equation, which
makes the calculations much more cumbersome and computer intensive.

A further simplification in the self-consistency calculation is that
we assume a quadratic momentum dependence for the single-particle
potential $U(p)$. This means that we can define an effective baryon
mass $M^*$ in terms of which the single-particle energy can be written as
\begin{mathletters}
\begin{eqnarray}
   E_i(p) &=& M_i + \frac{p^2}{2M_i} + {\cal R}e\,U_i(p) \label{stara}\\
          &\approx& M_i + \frac{p^2}{2M^*_i} + {\cal R}e\,U_i(0),
                                                         \label{starb}
\end{eqnarray}
\end{mathletters}
where $U_i(0)$ is easily obtained from Eq.~(\ref{sppot}), while $M^*_i$
is obtained from
\begin{equation}
   \frac{M^*_i}{M_i} = \left[1+\frac{{\cal R}e\,U_i(p_F)-{\cal R}e\,U_i(0)}
                 {p_F^2/2M_i}\right]^{-1}.
\end{equation}
The advantage of using Eq.~(\ref{starb}) is that the self-consistency
condition only needs to be calculated at $p=0$ and $p=p_F$, rather than
at a range of momentum values $0\leq p\leq p_F$, as required when using
Eq.~(\ref{stara}). Also, the binding energy does not require a numerical
integration, but is easily done analytically.
We checked for various cases that the parameterization of Eq.~(\ref{starb})
indeed fairly accurately represents the single-particle energy as obtained
from an explicit calculation using Eq.~(\ref{stara}).

In Eq.~(\ref{BGequ}), the Pauli operator needs to be expressed in terms
of ${\bf P}$ and ${\bf q}$.
Clearly, $|\mu_1{\bf P}\!+\!{\bf q}|\geq k_F^{(1)}$ and
$|\mu_2{\bf P}\!-\!{\bf q}|\geq k_F^{(2)}$ are both satisfied when
$q\geq\mu_1P\!+\!k_F^{(1)}$ and $q\geq\mu_2P\!+\!k_F^{(2)}$. Similarly,
when $q^2<k_F^{(1)2}\!-\!(\mu_1P)^2$ or $q^2<k_F^{(2)2}\!-\!(\mu_2P)^2$
at least one of the inequalities is not satisfied. For values of $q$
between these two limits, there are two restrictions on the angle
$\theta({\bf P},{\bf q})$, namely
\begin{eqnarray}
   \cos\theta &>& -\cos\theta_1 \equiv
             -\frac{(\mu_1P)^2+q^2-k_F^{(1)2}}{2\mu_1Pq}, \nonumber\\
   \cos\theta &<& \cos\theta_2 \equiv
              \frac{(\mu_2P)^2+q^2-k_F^{(2)2}}{2\mu_2Pq}.
\end{eqnarray}
Since the angle $\theta({\bf P},{\bf q})$ is integrated over, we can
approximate this latter constraint by taking an average value for the
Pauli operator $Q$. We therefore define
\begin{eqnarray}
   Q(q,P) &=& 1, \ \ \ \ \ \ \ {\rm if} \ \
          q\geq{\rm max}[\mu_1P+k_F^{(1)},\mu_2P+k_F^{(2)}] \nonumber\\
          &=& 0, \ \ \ \ \ \ \ {\rm if} \ \
          q^2<{\rm max}[k_F^{(1)2}\!-\!(\mu_1P)^2,
                        k_F^{(2)2}\!-\!(\mu_2P)^2]   \nonumber\\
          &=& {\rm min}[\cos\theta_1,\cos\theta_2], \ \ \ \ {\rm otherwise}.
\end{eqnarray}

In the partial-wave projection, the Bethe-Goldstone equation for
a system with isospin $T$ and total angular momentum $J$ becomes
\begin{eqnarray}
   G^{JT}_{l's',ls}(q',q;P,\omega)&=&V^{JT}_{l's',ls}(q',q)+\frac{2}{\pi}
         \sum_{l''s''}\int\!\!dq''\,q^{\prime\prime2} \nonumber\\
      && \times V^{JT}_{l's',l''s''}(q',q'') \frac{Q(q'',P)}
      {\omega\!-\!M\!-\!P^2/2M\!-\!q^{\prime\prime2}/2\mu\!-\!X+i\varepsilon}
         G^{JT}_{l''s'',ls}(q'',q;P,\omega),
\end{eqnarray}
where $X\!=\!0$ for the gap choice and
$X\!=\!U(|\mu_1{\bf P}\!+\!{\bf q}|)\!+\!U(|\mu_2{\bf P}\!-\!{\bf q}|)$
plus angle-averaging for the continuous choice.
The single-particle potential is obtained self-consistently from
\begin{equation}
   U({\bf p}_1)=\sum_{T,J,l,s}\frac{(2J+1)(2T+1)}{(2s_1+1)(2t_1+1)}
      2\pi\!\int_{-1}^{+1}\!\!d\cos\theta\int_0^{k_F^{(2)}}\!
      \frac{p_2^2dp_2}{(2\pi)^3}
           4\pi G^{JT}_{ls,ls}[k,k;P,E_1(p_1)+E_2(p_2)]_{AS}, \label{sppotpw}
\end{equation}
where we have explicitly separated off the angle dependence of the
$d^3p_2$ integral. The isospin factors are present to account for all
the contributions of the possible isospin states. (Our calculations are
done on the isospin basis.) Finally, the subscript $AS$ denotes that
we have to include both direct and exchanged (Hartree and Fock)
contributions. For identical particles, this can be accounted for by
multiplying the $G$ matrix from the Bethe-Goldstone equation with the
factor $1-(-1)^{l+s+\bar{t}}$, with $\bar{t}=1$ for singlet-even and
triplet-odd partial waves and $\bar{t}=0$ for singlet-odd and
triplet-even partial waves. (Note that $\bar{t}$ is equivalent to
the isospin in the case of pure $N\!N$ or pure $\Xi\Xi$ systems.)

If we now want to include all species of the octet baryons, the above
expressions can be easily generalized. First, the internal sum over
$l'',s''$ in the Bethe-Goldstone equation then also involves a sum over
all possible channels allowed for a particular two-baryon interaction.
Of course, the propagator needs to be modified to account for the
relevant masses and thresholds in each particular channel, and the
Pauli operator should contain the Fermi momenta belonging to the
relevant species.
Second, the single-particle potentials have to be summed over all
baryon species. Using the notation $U^{(B')}_B$ for the single-particle
potential of particle $B$ due to the interactions with particles $B'$
in the medium, and a bra-ket notation for the final-initial state
particles, we have
\begin{equation}
    U_B({\bf p}_1) = \sum_{B'} U_B^{(B')}({\bf p}_1),
\end{equation}
where
\begin{eqnarray}
    U_B^{(B')}({\bf p}_1) &=& \sum_{T,J,l,s}\frac{(2J+1)(2T+1)}
                                                 {(2s_B+1)(2t_B+1)}
               2\pi\int_{-1}^{+1}\!\!d\cos\theta\int^{k_F^{(B')}}\!\!
               \frac{p_2^2dp_2}{(2\pi)^3}        \nonumber\\
    && \hspace{2cm}\times\ 4\pi\langle BB'|G^{JT}_{ls,ls}[k,k;P,
                           E_1^{(B)}(p_1)+E_2^{(B')}(p_2)]_{AS}|
                           BB'\rangle.          \label{Uab}
\end{eqnarray}
The allowed values of $T,J,l,s$ depend on what particular baryons make
up the scattering process $B+B'\rightarrow B+B'$. For example,
$U_\Sigma^{(N)}$ gets contributions from direct isospin-3/2 $\Sigma N$
scattering, but also from the coupled-channel isospin-1/2
($\Lambda N,\Sigma N$) scattering. In our calculations
we include all partial waves up to $J=4$.
Finally, the binding energy per baryon is obtained from
\begin{eqnarray}
    \frac{E}{A}\rho &=& \sum_B (2s_B+1)(2t_B+1)\int^{(k^{(B)}_F)}
        \!\!\frac{d^3k}{(2\pi)^3}\left[\frac{k^2}{2M_B}+\frac{1}{2}
                          U_B(k)\right]  \nonumber\\
    &=& \frac{1}{2}\sum_B \rho_B \left[ {\cal R}e\,U_B(0)
        +\frac{3}{10}\left(\frac{1}{M_B^*}+\frac{1}{M_B}\right)
         k_F^{(B)2}\right].    \label{bind}
\end{eqnarray}

\section{RESULTS}
\label{sec:results}
\subsection{Pure systems}
\label{subsec:pure}
The first interesting question to ask is whether all of the isospin
symmetric matter ($T_z=0$) made of only one kind of hadrons is bound.
Our results are displayed in Fig.~\ref{strange:pureBB} for models
NSC97e (dashed curves) and NSC97f (solid curves).

The saturation curves for the purely nuclear system are very similar
to what is obtained for other one-boson-exchange $N\!N$ potentials
found in the literature. The results for the two NSC97 models are
practically indistinguishable, which, in fact, is true for all six
NSC97 models. This is a reflection of the fact that these models
all describe the $N\!N$ scattering data equally well.
 
We see that the pure $\Lambda$ system is not bound at all for both
models. The pure $\Lambda$ system was calculated including the coupling
to the $\Xi N$ and $\Sigma\Sigma$ channels. If this coupling is switched
off (and so only elastic $\Lambda\Lambda$ scattering is possible), the
curves shift to slightly higher values: about 10\% higher for NSC97e
and about 5\% higher for NSC97f.

The pure $\Xi$ system is more bound and saturates at higher densities
than the nucleon system. The difference in the results for NSC97e and
NSC97f is due to the fact that the $^3S_1$ partial-wave contribution to
the single-particle potential (which is large and positive) for NSC97f
is almost 40\% larger than for NSC97e. In addition, the $^3P_2$
partial-wave contribution (which is also large, but negative) for NSC97f
is about 10\% less attractive and largely compensates for an increased
attraction in the $^1S_0$ partial-wave contribution.
An important part of the attraction is due to the scalar-exchange part
of the potential. Since the existence of a nonet of scalar mesons with
masses below 1 GeV/$c$ is still highly controversial (especially the
low-mass isoscalar $\sigma$ meson), a comment is in order.
We first note that within a one-boson-exchange model for the $N\!N$
interaction, the scalar-exchange contribution plays a crucial role in
providing the required attraction. Whether this contribution represents
a true exchange of scalar mesons or just an effective parameterization
of two-pion exchange and more complicated interactions is a question
which goes beyond the scope of this paper.
Here we only want to mention that arbitrarily removing the $\sigma$
(or, in our case~\cite{Rij99}, the broad $\varepsilon$) contribution
renders purely nuclear matter unbound at all densities. However,
the pure $\Xi$ system still remains bound, although it saturates at
a smaller density.

Pure $\Sigma$ matter with $T_z=0$ is expected to be highly unstable,
because it can decay strongly into $\Lambda$ matter via
$\Sigma^+\Sigma^-\rightarrow\Lambda\Lambda$ and
$\Sigma^0\Sigma^0\rightarrow\Lambda\Lambda$. However, we can consider
pure $\Sigma^+$ or pure $\Sigma^-$ matter with $T_z=\pm2$, respectively.
Inclusion of the Coulomb interaction in this charged system is beyond
our present calculation (and is likely to modify the result), and so
the third panel of Fig.~\ref{strange:pureBB} represents the result
without the Coulomb interaction and is included for illustrative
purposes only. The difference in the results for NSC97e and NSC97f
is due to the fact that the single-particle potential for NSC97f is
slightly more attractive for almost all partial-wave contributions,
which adds up to a substantial difference in the total single-particle
potential.

The results of these calculations show that for the employed baryon-baryon
models of Ref.~\cite{Rij99,Sto99} the only possible long-lived strange
pure system within SU(3) is the $\Xi$ system, which is stable against
strong decays. This suggests that the inner core of high-density
astrophysical objects could be rich in $\Xi$ particles.

\subsection{$\protect\bbox{\{N,\Lambda\}}$ systems}
\label{subsec:NL}
We next consider the change of the nuclear binding by adding $\Lambda$'s.
The results are shown in Fig.~\ref{strange:LfracN}. We should point
out that, even though the $\Sigma$ and $\Xi$ are not explicitly
included as part of the medium, the coupling to these particles
via the transition potentials $V(\Lambda N\rightarrow\Sigma N)$ and
$V(\Lambda\Lambda\rightarrow\Xi N,\Sigma\Sigma)$ {\it are\/} included.
We see that the binding first slightly increases until the $\Lambda$
fraction reaches about 10\%. The system rapidly becomes less bound
when more $\Lambda$ particles are added.

The most stable system occurs at about $\rho=0.28$ fm$^{-3}$ and
$\chi_\Lambda=0.1$ with $E/A_B=-13.2$ MeV for NSC97e and at about
$\rho=0.27$ fm$^{-3}$ and $\chi_\Lambda=0.05$ with $E/A_B=-12.7$ MeV
for NSC97f. The $\Lambda$ fraction at the minimum is very similar to
the $\chi_\Lambda\approx0.06$ found in Ref.~\cite{Sch98}, but in their
case the minimum occurs at a lower density with a higher binding energy
($\rho=0.23$ fm$^{-3}$ with $E/A_B=-16.0$ MeV). This difference is due
to various reasons.
First of all, the larger binding energy found in Ref.~\cite{Sch98}
is due to the fact that they use the continuous choice for the
single-particle energy, whereas we use the gap choice. The smaller
density at which their saturation occurs is due to the different
$N\!N$ potential that is employed: they use the parameterized Paris
$N\!N$ potential~\cite{Lac80}, whereas here we use the $N\!N$
potential as given by the NSC97 models.
Furthermore, the results of Ref.~\cite{Sch98} do not include the
effect of $\Lambda$'s interacting with themselves. We {\it do\/}
include the $\Lambda\Lambda$ interaction and, although we find
that the contribution of $U_\Lambda^{(\Lambda)}$ to the total
single-particle potential is rather small, its influence starts
to become noticeable as the $\Lambda$ density increases.
At low $\Lambda$ density the individual partial-wave contributions
to $U_\Lambda^{(\Lambda)}$ are almost all negative, but as the
$\Lambda$ density increases some of them start to give (relatively
important) positive contributions. As a consequence, the effect of
the inclusion of the $\Lambda\Lambda$ interaction for increasing
$\chi_\Lambda$ is to shift the density at which saturation occurs
to lower values. The effect is most pronounced for NSC97f.
In Ref.~\cite{Sch98} the saturation density is more or less independent
of the $\Lambda$ fraction; see their Fig.~5.

Our results suggest that multi-$\Lambda$ systems produced in
relativistic heavy-ion collisions through, for example, the coalescent
mechanism could be loosely bound. However, the presence of a large
fraction of $\Lambda$'s in the inner core of a dense star seems
unlikely, since a too large fraction tends to destabilize it.

\subsection{$\protect\bbox{\{N,\Lambda,\Xi\}}$ systems}
\label{subsec:NLX}
A further difference between the present work and that of
Ref.~\cite{Sch98} is that here we can also include the $\Xi$ (and
$\Sigma$) as part of the medium.
To investigate the influence of including $\Xi$'s, we perform
calculations for a system consisting of $N$, $\Lambda$, and $\Xi$.
The $\Sigma$'s are excluded since they can easily be annihilated,
as stated before. Another motivation for their exclusion is that the
Q-values for the strong transitions $\Sigma N\rightarrow\Lambda N$,
$\Sigma\Sigma\rightarrow\Lambda\Lambda$, $\Sigma\Lambda\rightarrow\Xi N$,
and $\Sigma\Xi\rightarrow\Lambda\Xi$ are about 78, 156, 50, and 80 MeV,
respectively. To Pauli block these processes, we need a rather high
density of $\Lambda$.
On the other hand, the Q-value of $\Xi N\rightarrow\Lambda\Lambda$
is only about 28 MeV. The presence of $\Xi$ could then help prevent
the collapse of the $\{N,\Lambda,\Xi\}$ system since the
$\Lambda\Lambda\rightarrow\Xi N$ reaction can be Pauli blocked.
The importance of including $\Xi$'s was first pointed out in
Ref.~\cite{Sch93}.

It is interesting to first investigate the charge-neutral systems
consisting of only $N$, $\Lambda$, and $\Xi$. They can be formed with
a density distribution of $\rho_p=\rho_n=\rho_{\Xi^0}=\rho_{\Xi^-}$.
Our results are shown in Fig.~\ref{strange:LfracNX}. We see that
the systems are only loosely bound. The $\Lambda$ density is too
low to Pauli block the $\Xi N\rightarrow\Lambda\Lambda$ process.
This suggests that a charge-neutral strangeness-rich system is
unlikely to be seen in nature or to be created in relativistic
heavy-ion collisions.

We now turn to investigating the dependence of the binding energy
of the ${N,\Lambda,\Xi}$ system on the strangeness per baryon.
As discussed in previous works~\cite{Sch93,Sch94,Gal95}, this
dependence is most relevant to the investigation of relativistic
heavy-ion collisions. For this purpose, it is useful to define the
fractions $\chi_i=\rho_i/\rho$ for the different species
$i=N,\Lambda,\Xi$, as in Eq.~(\ref{rhotot}).
The strangeness per baryon for the $\{N,\Lambda,\Xi\}$ systems can
then easily be calculated as $f_s=\chi_\Lambda+2\chi_\Xi$.
Since we work on the isospin basis, we will only consider systems
with $\rho_n=\rho_p=\frac{1}{2}\rho_N$ and
$\rho_{\Xi^0}=\rho_{\Xi^-}=\frac{1}{2}\rho_\Xi$.
The charge per baryon is then $f_q=\frac{1}{2}(\chi_N-\chi_\Xi)$.
For each $\{N,\Lambda,\Xi\}$ system with a given $\chi_N$ we
carried out calculations for various combinations of
$(\chi_\Lambda+\chi_\Xi)=(1-\chi_N)$.
In general, we find that the system is less bound for high $\chi_\Xi$,
as could already be inferred from comparing Figs.~\ref{strange:LfracN}
and \ref{strange:LfracNX}.
The reason is that $U_\Xi^{(\Xi)}$ becomes more negative (attractive)
as $\chi_\Xi$ increases, but this is compensated by positive (repulsive)
contributions from $U_{N,\Lambda}^{(\Xi)}$ and $U_\Xi^{(N,\Lambda)}$.
The cancellations are large enough to prevent $\rho_\Xi$ from becoming
too large, and so the large binding energies as found close to the
saturation point of the pure $\Xi$ system (see Fig.~\ref{strange:pureBB})
cannot be reached.

These calculations allow us to examine the $f_s$ ($f_q$) dependence of
the binding energies at the saturation point of an $\{N,\Lambda,\Xi\}$
system. The results are shown in Fig.~\ref{strange:NLXmix}. In each
case, the curves cover the allowed $f_s$ values that can be reached.
We see that as $f_s$ (and $f_q$) increases, the system becomes less
bound. When $\chi_N=0.7$, the purely $\{N,\Lambda\}$ system (i.e.,
$\chi_\Xi=0$) has the lowest binding, but when $\chi_N$ gets smaller
than 0.6, systems with increasing $\chi_\Xi$ are preferred. This
follows from the fact that in those cases the curves show a (shallow)
minimum. However, when $\chi_N\lesssim0.4$ the system becomes unbound.
Our results contradict the results from the relativistic mean-field
calculations of Refs.~\cite{Sch93,Sch94,Gal95} (see Figs.~2 and 3 of
Ref.~\cite{Sch93}).
The differences, of course, could be due to finite-size effects, since
these authors consider a shell model for finite $\{N,\Lambda,\Xi\}$
systems (up to very large $A_N=310$).
But it is more likely that the difference is due to the differences
in dynamical content of the calculations. This can be understood
by observing that the mean-field calculation within the
Brueckner-Hartree-Fock approach amounts to neglecting the residual
baryon-baryon interaction terms in calculating $E/A_B$.
Namely, the mean-field results can be obtained from Eq.~(\ref{bind})
by making the change $\frac{1}{2}U_B(k)\rightarrow U_B(k)$ in
Eq.~(\ref{bind}). In all of the cases, $U_B(k)$ is comparable to
kinetic energies and, hence, the finite-size mean-field results can
be radically different from our full calculations.

\subsection{Inclusion of $\protect\bbox{\Sigma}$}
\label{subsec:NLSX}
Although the $\Sigma$ particles can be easily annihilated, as stated
before, their presence largely increases the binding of the system,
which might be of relevance in the formation and stability of such a
system.
To demonstrate the effect of adding $\Sigma$'s we first consider a
charge-neutral $\{N,\Lambda,\Xi\}$ system with $\chi_N=\chi_\Xi$,
and so $f_s=1$. For each choice of $\chi_N=\chi_\Xi$ we can add
different fractions of $\Sigma$'s as long as
$(\chi_\Lambda+\chi_\Sigma)=(1-\chi_N-\chi_\Xi)$. Note that these
systems all still have $f_s=1$.
In order to prevent having to show numerous figures or tables, we
will here only consider the systems with $\chi_\Lambda=\chi_\Sigma$.
The results are shown in Fig.~\ref{strange:Sigma}. They should be
compared with the saturation minima in Fig.~\ref{strange:LfracNX}.
We clearly see that adding also $\Sigma$'s drastically improves the
binding of the system compared to adding only $\Lambda$'s.

The increase in binding due to adding $\Sigma$'s allows us to further
explore the $f_s$ dependence in the region beyond $f_s<1$. Again,
numerous choices for the different particle fractions are possible,
but here we will only restrict ourselves to systems with equal fractions
for the strange particles, i.e., $\chi_Y\equiv\chi_\Lambda=\chi_\Sigma=
\chi_\Xi$.
We find that these systems are bound for any value of $\chi_Y$.
The $f_s$ dependence of this mixed system is shown in
Fig.~\ref{strange:SperB}. We see that as $f_s$ increases, the system
becomes more bound, except in the low $f_s$ region.
Note that at the highest $f_s=2$, the system is the pure $\Xi$ system
as in Fig.~\ref{strange:pureBB}. Of course, our special choice for
$\chi_Y$ means that the strangeness per baryon can only go up to
$f_s=4/3$, and so the result for $f_s>4/3$ in Fig.~\ref{strange:SperB}
is not calculated, but is rather obtained by simply extrapolating to
the pure $\Xi$ result which has $f_s=2$.
Comparing with the results of Fig.~\ref{strange:NLXmix} for the
$\{N,\Lambda,\Xi\}$ system, it is clear that the addition of the
$\Sigma$ component drastically changes the $f_s$ dependence, but it
is important to remember here that the system containing $\Sigma$ is
highly unstable against strong decays, as discussed above.
The results shown in Fig.~\ref{strange:SperB} perhaps cannot be
verified in relativistic heavy-ion collisions. However, it could
represent the situation in the core of neutron stars in which the
presence of high density $e^-$ can produce a lot of $\Sigma^-$,
initiated by the reaction $e^-+n \rightarrow \Sigma^-+\nu$.

\section{CONCLUSION}
\label{sec:conc}
We have investigated strange hadron matter in the context of a
baryon-baryon potential model based on SU(3) symmetry. The parameters
of the potential model were fitted to the $N\!N$ and $Y\!N$ scattering
data, and the assumption of (broken) SU(3) symmetry allows us to
extend the model to also describe the other interaction channels
that are allowed for the baryon octet; i.e., the $YY$, $\Xi N$,
$\Xi Y$, and $\Xi\Xi$ interactions. The potential for these
interactions is defined without the necessity of having to introduce
new free parameters.

The calculations have been carried out by using the
Brueckner-Hartree-Fock approximation. Within the framework that we use
to define the potential model for the baryon-baryon interactions, we
find that the pure $\Lambda$ system is unbound, whereas the pure $\Xi$
system is more strongly bound and saturates at higher densities than
the pure $N$ system.
Adding $\Lambda$'s to pure nuclear matter slightly increases binding,
as long as the $\Lambda$ fraction is less than about 10\%. Larger
fractions cause a decrease in binding. Adding $\Xi$'s, of importance
due to the reaction $\Lambda\Lambda\rightarrow\Xi N$, drastically
reduces binding, and so $\{N,\Lambda,\Xi\}$ systems are only (weakly)
bound for nucleon fractions larger than 40\%.
Our results represent a step forward, since the previous
Brueckner-Hartree-Fock calculation~\cite{Sch98} did not include
the $\Xi$'s.

By carrying out extensive calculations for the $\{N,\Lambda,\Xi\}$
and $\{N,\Lambda,\Sigma,\Xi\}$ systems, we have predicted the
dependence of the binding energy on the strangeness per baryon,
a quantity that is needed to be determined as precisely as possible 
for identifying the strange quark matter created in relativistic
heavy-ion collisions. Our results are significantly different from
previous calculations based on relativistic mean-field models.
We argue that the differences are mainly due to the two-body
correlations, which are neglected in relativistic mean-field models.

To close, we would like to point out that the present work is merely
a first step towards a rigorous many-body calculation. In the future,
we need to investigate the three-body terms in the Brueckner-Hartree-Fock
approach. This is particularly important if one is to give more
precise predictions in the higher-density regions. Another reason is
that perhaps it is only at that order that the issue of using the gap
choice or the continuous choice for the single-particle energies in
solving the self-consistent Eqs.~(\ref{BGequ}) and (\ref{sppot})
becomes irrelevant, as was shown in a recent work on ordinary
nuclear matter~\cite{Son98}.
Our investigation in this direction will be published elsewhere.

\acknowledgments
This work was partly supported by the U.S.\ Department of Energy,
Nuclear Physics Division, under Contract No.\ W-31-109-ENG-38.

\begin{figure}
\caption{$^1S_0$ phase shifts for elastic identical-particle
         scattering, for models NSC97e and NSC97f.}
\label{strange:phs1s0}      
\end{figure}

\begin{figure}
\caption{Saturation of pure systems. The dashed curve and solid
         curve are the predictions of model (e) and (f), respectively.
         The pure $\Sigma$ system represents the $T_z=\pm2$ case
         without the Coulomb interaction.}
\label{strange:pureBB}      
\end{figure}

\begin{figure}
\caption{Saturation of $\{N,\Lambda\}$ systems for various fractions
         $\rho_{\Lambda}/(\rho_N+\rho_\Lambda)$.}
\label{strange:LfracN}      
\end{figure}

\begin{figure}
\caption{Saturation of charge-neutral $\{N,\Lambda,\Xi\}$ systems for
         various fractions $\rho_{\Lambda}/(\rho_N+\rho_\Lambda+\rho_\Xi)$.}
\label{strange:LfracNX}      
\end{figure}

\begin{figure}
\caption{$f_s$ dependence of the $\{N,\Lambda,\Xi\}$ system
         for three given fractions of nucleons $\chi_N$.}
\label{strange:NLXmix}      
\end{figure}

\begin{figure}
\caption{Saturation of charge neutral $\{N,\Lambda,\Sigma,\Xi\}$
         system with $\chi_N=\chi_\Xi$ and $\chi_\Lambda=\chi_\Sigma$,
         as a function of $\chi_{\Lambda+\Sigma}$.}
\label{strange:Sigma}      
\end{figure}

\begin{figure}
\caption{$f_s$ dependence of the $\{N,\Lambda,\Sigma,\Xi\}$ system
         with $\chi_\Lambda=\chi_\Sigma=\chi_\Xi$.}
\label{strange:SperB}      
\end{figure}

\end{document}